\documentstyle[preprint,aps,epsfig]{revtex}
\newcommand{\bx}{{\bf x}}
\newcommand{\bv}{{\bf v}}
\newcommand{\bF}{{\bf F}}
\newcommand{\bG}{{\bf G}}
\newcommand{\bC}{{\bf C}}

\begin{document}
\title{Population dynamics advected by chaotic flows: \\ a discrete-time map
approach.}
\author{ Crist\'obal L\'opez$^{1}$,
Emilio Hern\'andez-Garc\'\i a$^{1}$,
 Oreste Piro$^{1}$,
Angelo Vulpiani$^{2}$ and Enrico Zambianchi$^{3}$
}
\address{$^1$ Instituto Mediterr\'aneo de Estudios Avanzados (IMEDEA),\\
E-07071 Palma de Mallorca, Spain. \\ $^2$ Dipartimento di Fisica,
Universit\'a di Roma `La Sapienza' and I.N.F.M., P.le A. Moro 2, I-00185, Roma,
Italy. \\ $^3$ Istituto di Meteorologia e Oceanografia, Istituto
Universitario Navale, \\ Via Acton 38, I-80133 Napoli, Italy. }

\maketitle
\today

\begin{abstract}

A discrete-time model of reacting evolving
fields, transported by a bidimensional chaotic fluid flow, is
studied. Our approach is based on the use of a Lagrangian scheme
where {\it fluid particles} are advected by a $2d$ symplectic map
possibly yielding Lagrangian chaos. Each {\it fluid particle}
carries concentrations of active substances which evolve according
to its own reaction  dynamics. This evolution is also
modeled in terms of maps. Motivated by the question, of relevance
in marine ecology, of how a localized distribution of nutrients or
preys affects the spatial structure of predators transported by a
fluid flow, we study a specific model in
which the population dynamics is given by a logistic map with
space-dependent coefficient, and advection is given by the
standard map. Fractal and random patterns in the Eulerian spatial
concentration of predators are obtained under different
conditions. Exploiting the analogies of this coupled-map
(advection plus reaction) system with a random map, some features
of these patterns are discussed.

\end{abstract}
\vspace{1.5cm}

\newpage

{\bf 
The spatial structure of passive fields transported by a
fluid flow is an important question in fluid and nonlinear dynamics.
Many important situations require in addition 
to take into account
chemical or biological interactions between the substances
transported by the flow. In particular, our motivation comes from
the general question on the mechanisms for plankton inhomogeneity
in the ocean, and more specially for the spatial patterns that
would be reached by plankton or marine hervibores grazing from a
localized source of nutrients. We present a general methodology,
based in coupled discrete-time dynamical systems describing
reaction or population dynamics and advection in the Lagrangian
framework, and apply it to a model of logistic population dynamics
in the presence of a localized source of nutrients and chaotic
advection. Fractal and random spatial features develop in the
concentration patterns, which can be understood in terms of an
analogy with random maps.
}

\section{INTRODUCTION}

Patchiness, or the uneven distribution of substances of organisms,
is ubiquitously observed in the ocean
\cite{abraham,fasham,steele,stehen}.
 In complex situations such as
the marine ecosystems, characterized by the interplay of
population dynamics and an ambient fluid motion which may
differently affect individual populations, the question of how a
localized availability of nutrients (or preys, or, in chemical
terms, activators) may affect the distribution of primary
producers (or predators, or inhibitors, respectively; in the
following we refer to
 preys and predators) is a crucial and challenging problem.

 According to \cite{fasham}, the possible causes originating
patchiness in marine ecosystems may be grouped in different
categories: fluid motion, biological growth coupled with
dispersive processes, less ubiquitous mechanisms like swarming,
vertical migration, and others.
In addition to patchiness, localized availability of {\it food}
may occur in correspondence of localized sub-ecosystems, such as
Posidonia Oceanica beds (for a review see \cite{mazella}): they
display a quite complex structure in which the most important
features arise from direct consumption of the plant and
epiphyte-herbivore interactions, although a portion of the trophic
chain is based on suspended matter \cite{mazella}.

A comprehensive description of such processes leads to the
study of the so called advection-reaction-diffusion
equations.
 These are partial differential equations of the type:
\begin{equation}
\frac{\partial C_i(\bx,t)}{\partial t}+(\bv(\bx,t)\cdot
\nabla )C_i =R_i(C_1,...,C_N,\bx,t)+D_i\triangle C_i.
 \label{ard}
\end{equation}
Where $C_i(\bx,t)$  ($i = 1,...,N$) is the concentration of the
$i$-th reactive (in biological or chemical terms) species or
substance, and the functions $R_i$ describe the reaction, or the
population, intrinsic dynamics. The possible explicit spatial and
temporal dependence may model the influence of temporal and
spatial inhomogeneities in food, temperature, etc.  The term
$(\bv(\bx,t)\cdot
\nabla )C_i$ represents advection by a given solenoidal (i.e.
incompressible $\nabla \cdot \bv=0$) velocity field $\bv(\bx,t)$.
Finally, the term $D_i\triangle C_i$ describes diffusion of the
$i$-th species or substance with diffusivity $D_i$. In writing
(\ref{ard}) we are assuming that the evolution of the advected
concentrations does not affect that of the underlying flow
$\bv(\bx,t)$. This is definitely reasonable at the scales we are
interested in, even though it is worth mentioning that at much
smaller scales the presence of organisms may affect the
rheological properties of seawater \cite{jb,jwm}.

In some situations it would be necessary to consider a different
velocity field for each of the $N$ concentrations $C_i$. This may
happen, for instance, if different organisms live at different
mean depths, leading to differences in the experienced flow. In
such cases one has to replace Eq.~(\ref{ard}) by
\begin{equation}
\frac{\partial C_i(\bx,t)}{\partial t}+(\bv_i(\bx,t)\cdot
\nabla )C_i =R_i(C_1,...,C_N,\bx,t)+D_i\triangle C_i.
 \label{ard2}
\end{equation}
In this paper, however, we will only consider the case given by
Eq.~(\ref{ard}) in which the same velocity field advects all the
substances.

Introducing the Lagrangian time derivative $\frac{d}{dt}=
\frac{\partial}{\partial t}+ \bv \cdot \nabla$,
Eq.~(\ref{ard}) can be written in the form
\begin{equation}
\frac{d C_i(\bx,t)}{dt}
=R_i(C_1,...,C_N,t)+D_i\triangle C_i.
\end{equation}
If diffusion is neglected, it is simple to write the $C_i(\bx,t)$
in terms of the solutions of the Lagrangian evolution equation
\begin{equation}
\frac{d \bx(t)}{dt}= \bv(\bx(t),t),
\label{velo}
\end{equation}
and the reactive evolution equation
\begin{equation}
\frac{dC_i(t)}{dt} =R_i(C_1,...,C_N,\bx(t),t).
\label{reac}
\end{equation}
This set of coupled ordinary differential equations describe the
advection-reaction process in a Lagrangian frame: fluid particles
move according to (\ref{velo}), and reactions among the $C_i$'s
occur inside each fluid particle, as expressed by (\ref{reac}),
where $C_i(t)$ are the concentrations at a particular fluid
particle (the one at $x(t)$ at time $t$, i.e. $C_i(t) \equiv
C_i(\bx(t),t)$). Denoting by $S^t$ and $L^t$ the formal solutions
of (\ref{velo}) and (\ref{reac}) respectively (i.e., $\bx(t)=S^t
\bx(0)$ and $\bC(t)=L^t \bC(0)$, with $\bC=(C_1,C_2,...,C_N)$) we
can write the solution of Eq.~(\ref{ard}) in the form (with
$D_i=0$)
\begin{equation}
\bC(\bx,t)=L^t \bC(S^{-t}\bx,0).
\label{cosol}
\end{equation}
The case $D_i \ne 0$ needs a more elaborated treatment. In
addition to the time dependence, $L^t$ will have also an explicit
space dependence if the $R_i$'s have it.

 Obviously, the
detailed understanding of the above class of partial differential
equations constitute a formidable task. However, at this stage of
development, we are just interested in the search for generic
behaviors expected when few typical characteristics of the flow
and of the population dynamics are considered. Thus, if, for
example, we concentrate on flows of geophysical nature, horizontal
motion turns out to be much more intense than vertical one as soon
as one considers scales larger than a few kilometers. This
justifies restricting in the following to incompressible
twodimensional flows. A turbulent bidimensional flow would be a
way to model the irregular advection process to which suspended
matter is subjected in real oceans. There are however simpler
classes of flows which share some basic characteristics with
turbulence, but are much more accessible to analysis: Lagrangian
chaotic flows\cite{Ottino,CFPrev}. These are smooth velocity
fields, with some simple time dependence in the Eulerian
description, but which lead to chaotic trajectories of fluid
elements, with the associated stretching and folding, in the
Lagrangian description. In this restricted framework, it is well
known that even periodic time-dependence in two-dimensional
incompressible flows leads generically to chaotic motion of fluid
particles.

Rather than integrating the full equations describing the
continuous in time dynamics, and since our interest lies mainly in
a qualitative characterization of the population system, we will
resort in this paper to a discrete in time mapping-approach in
terms of discrete-time dynamical systems. This approach is
numerically very efficient, and has proven to be extremely
productive to study the impact of chaotic advection on mixing
\cite{Ottino,CFPrev,CFPPhysD}. The main idea is to mimic the
advection and reaction processes in terms of maps that capture the
main features of each aspect. Thus, since we will be looking at
processes taking place in 2d incompressible flows, the advective
part of our model is naturally described by a twodimensional
symplectic map. It is well known that Lagrangian motion in such
systems is typically chaotic and with a rather rich behavior. In
addition, the transported fluid parcel contains concentrations of
active chemical substances or biological specimens subjected to a
specific dynamics that will be also modeled in terms of a map.
Also, it is worth noticing that even though this is not done in
this work, diffusion can be easily reincorporated into the model
by averaging, after each iteration of the maps, the concentration
of the different species contained in the fluid elements over a
region of size $l\sim\sqrt{D_i\tau}$, where $D_i$ is the
corresponding diffusivity and $\tau$ is a characteristic time
scale of the system.

The general ideas and formalism sketched above will be made more
concrete in the following, and applied to tackle our main problem:
the {\it influence of inhomogeneities of the distribution of preys
on that of predators}. We will make use of some known results for
random maps, and compare our discrete-time approach with results
obtained in a continuous-time description of the problem
\cite{filam,filam2}. In Section II we will discuss the approach to
population dynamics in terms of maps; in Section III our
particular model is presented and compared with results for a
random logistic map; the analogy helps in the interpretation of
the resulting spatial structure of predators, which is described in Section
IV, and discussed in Section V.

\section{A DISCRETE-TIME APPROACH}

Let us now present the general idea of our approach for the
analysis of the population dynamics in terms of maps.

It is easy to understand that for time-periodic velocity fields,
i.e. $\bv(\bx,t)=\bv(\bx,t+T)$ where $T$ is the period,
Eq~(\ref{velo}) can be described by a discrete-time dynamical
system. The position $\bx(t+T)$ is univocally determined by
$\bx(t)$. In addition (because of the periodic velocity field) the
map $\bx(t) \to \bx(t+T)$ cannot depend on $t$. Since a periodic
time dependence is enough to induce Lagrangian chaos, and because
of the above mathematical simplifications, we particularize our
study to time-periodic velocity fields, for which we can write
\begin{equation}
\bx(t+1)=\bF(\bx(t)),
\label{sol}
\end{equation}
Now, time is measured in units of the period $T$. If $\bv$ is
incompressible, the map (\ref{sol}) is volume (area in $2d$)
preserving, i.e., $\left| \det (\frac{\partial F_i}{\partial x_j})
\right|=1$. In $2d$ the map (\ref{sol}) is symplectic, i.e. the
discrete-time version of a Hamiltonian system. Usually, it is not
simple at all to obtain $\bF(\bx(t))$ for a given $\bv(\bx,t)$.
However, one can directly write models for $\bF$ which contain the
qualitative features of the flow one is trying to model.

In addition, the transported fluid parcel contains species
subjected to their own population dynamics. Denoting the solution
of (\ref{reac}) after one period of the flow ($L^T$) by $\bG$, the
evolution rule for the interacting concentrations ${\bf
C}=(C_1,C_2,...C_n)$ is expressed in terms of a map:
\begin{equation}
\bC(t+1)=\bG(\bC(t))
\end{equation}

As before, $\bG$ will carry additional explicit time and space
dependencies if (\ref{reac}) is not autonomous in space or time.
The discrete-time version of Eq.~(\ref{cosol}) is:
\begin{equation}
\bC(\bF(\bx),t+1)=\bG(\bC(\bx,t)).
\label{cosolDis}
\end{equation}

In the following we particularize this general approach to a
particular model.

\section{A PARTICULAR MODEL AND ITS RELATIONSHIP WITH A RANDOM
LOGISTIC MAP}

The main interest of our study is to consider the problem of  how
the spatial structure of the prey spatial distribution may affect
the one of the predators. In this section we study a particular
model and present its analogies with the random map. This will
enable us to use the already known properties of this random map
to get a further insight into the influence of the distribution of
preys on the predator patterns. We will considered a
single-species population dynamics, i.e. the predator evolves for
fixed prey distribution, and under the influence of the flow, but
the distribution of the prey is a non-dynamic variable, in the
sense that it is not transported by the flow and is maintained at
fixed values undisturbed by the predator action. This is the
simplest setting in which the effects of a localized source of
nutrients on an advected predator will show up.

The model is the following: the positions of the {\it fluid
parcels} are advected by a standard map \cite{Lichtenberg}, i.e. a
2d symplectic map defined in the square of side $2\pi$ by
\begin{eqnarray}
x(t+1) & = & (x(t) + K \sin  y(t+1))  \ {\rm mod} \ 2\pi \\ y(t+1)
& = & (y(t) + x(t) ) \ {\rm mod} \ 2\pi
\label{stmp}
\end{eqnarray}

It is not integrable for $K \ne 0$. As $K$ increases chaotic
regions occupy larger areas, and the original KAM tori (regular
non-chaotic orbits) are successively destroyed. For $K $ large
enough the KAM tori occupy a very small region and practically the
whole phase space  is a unique chaotic region.

The model is completed by stating the evolution rules for the
predator-prey interactions. We denote with  $n(\bx)$  the
stationary spatially inhomogenous distribution of preys, and with
$C(\bx,t)$ the concentration of predators in point $\bx$ at time
$t$. We take it to evolve in each fluid parcel according to the
well known logistic map : $C (t+1)=G(C)= r C(t)(1-C(t))$, but with
a growth rate parameter $r$ determined by the presence of preys,
i.e., $r=\mu n$. The complete evolution equation (\ref{cosolDis})
is
\begin{equation}
C (\bF (\bx),t+1)=\mu n(\bx) C (\bx,t) (1-C (\bx,t)).
\label{lagran}
\end{equation}
The standard map has been written in the compact form (\ref{sol})
with $\bx=(x,y)$.

We now introduce the particular form  of the localized prey
distribution $n(\bx)$:
\begin{equation}
r=\mu n(\bx) = \left\lbrace
\begin{array}{c l}
 r_1  & \text{if  ${x} \in [\pi(1-p),\pi(1+p)]$}, \ \forall y\\
        r_0 & \text{otherwise}.
\end{array}
\right.
\label{step}
\end{equation}
with $0 \leq p \leq 1$. We are suggesting a striped spatial
distribution of the preys (with strip width $2 \pi p$), which
basically represents (due to the $2\pi$-periodicity of the flow)
the simplest, space-periodic fashion to mimic a patchy
distribution. A fraction $p$ of system area has the value $r=r_1$,
and fraction $1-p$, the $r=r_0$. The heuristic idea that will
guide our analysis is that, if mixing provided by the advection
map is strong enough, fluid parcels will visit regions with the
different values of $r$ in a stochastic way, so that the
Lagrangian evolution of the concentrations will be well described
by a random logistic map, i.e. a map of the form
\begin{equation}
C(t+1)=a_t C(t)\left( 1-C(t) \right),
\label{randommap1}
\end{equation}
where the random variable $a_t$ can take only two values
\begin{equation} a_t = \left\lbrace
\begin{array}{c l}
 r_0 & \text{with probability $1-p$}, \\
 r_1 & \text{with probability $p$},
\label{randommap2}
\end{array}
\right.
\end{equation}
and $a_{t+1}$ is independent of any previous $a_t$.

The random map (\ref{randommap1}) has been studied in
\cite{vulpio} for $r_0=1/2$ and $r_1=4$. This corresponds to the
situation in which for a value of $r=r_0$ the population dynamics
is attracted by a fixed point, whereas chaotic population dynamics
occurs for $r=r_1$. The alternancy in time of these two tendencies
gives rise to nontrivial behavior. Numerical (and some analytical)
computations \cite{vulpio} give the following results for
(\ref{randommap1}) with $r_0=1/2$ and $r_1=4$, for different
values of $p$:

\begin{itemize}
  \item[i)] If $p \leq p_1= 1/3$ the Lyapunov exponent
$\lambda=\lim_{N\to \infty}
\frac{1}{N}\sum_{i=0}^{N-1}\ln\left|a_t\left(1-2C(t)\right)\right|$
 is negative, i.e.  there is exponential convergence of two initially close
sequences $C(t)$ generated with the same sequence ${a_t}$ but
slightly different initial conditions $C(0)$. In this case, the
sequences are attracted by $C=0$.
  \item[ii)]If $p_1 \leq p \leq p_2 \simeq 0.5$ the Lyapunov
  exponent is negative again, but now the sequences do not converge
to any fixed point. They wander in an irregular and seemingly
chaotic manner (of the {\sl on-off intermittency} type). The
meaning of the negative value of $\lambda$ is that close initial
values of $C$ evolve, under the same sequence $a_t$, towards the
same irregular trajectory. This is a case of chaotic
synchronization related to the phenomenon of synchronization by
noise \cite{Lai,UPON}.
  \item[iii)] If $p \geq p_2$ the Lyapunov exponent is positive,
  i.e.  there exponential divergence of two initially close
sequences $C(t)$, behaving both chaotically.
\end{itemize}

\begin{figure}
\begin{center}
Figure 1 \\
L\'opez et al.
\end{center}
\vspace{2cm}
\centering
\epsfig{file=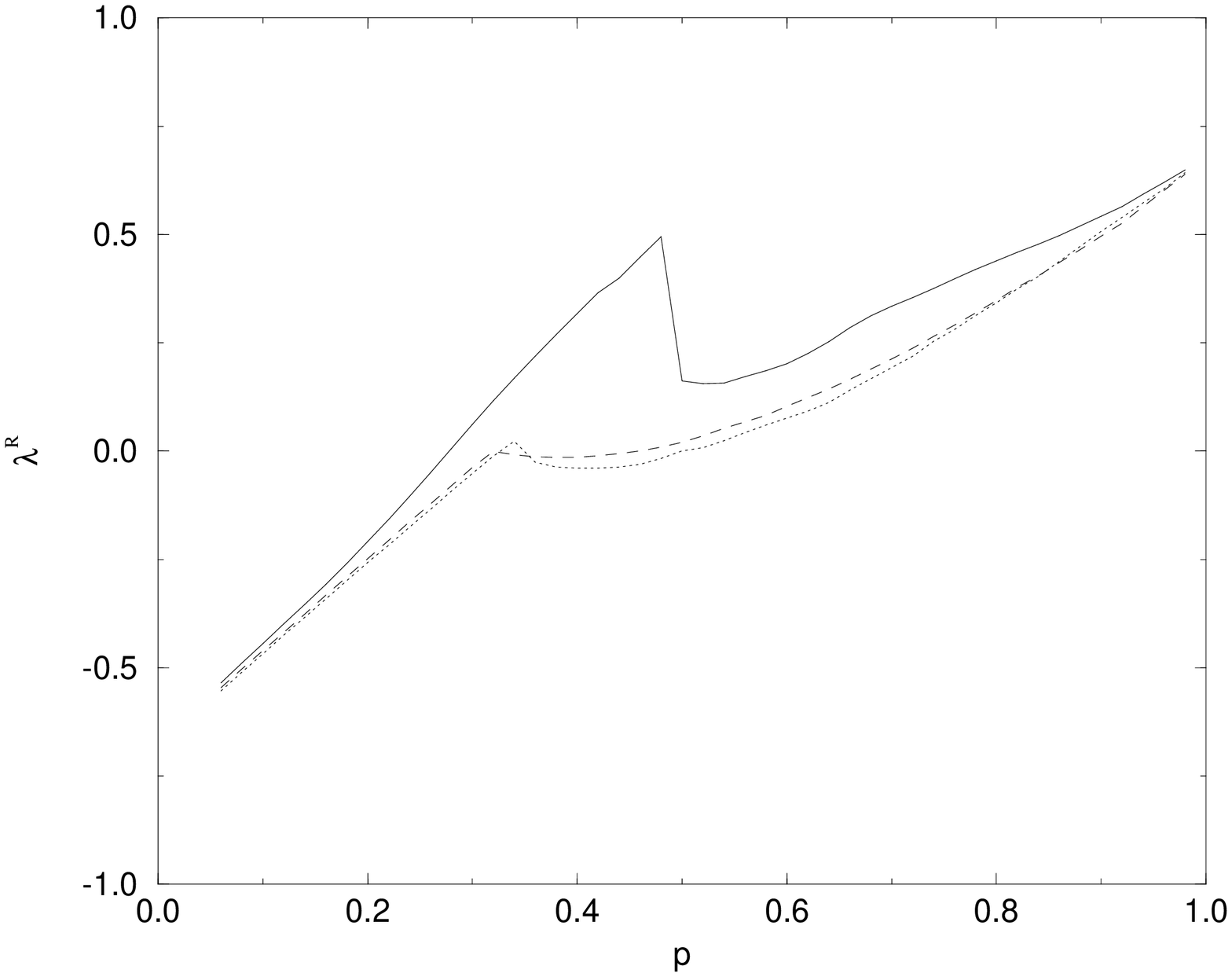,width=10cm, angle=-90}
 \caption{ Reaction Lyapunov exponent, $\lambda^R$, calculated for
different values of $K$.
 Solid line corresponds to $K=1.5$,
dotted line to $K=9$ and dashed line to $K=19$.
}
\label{fig:lambdar}
\end{figure}

As a first check confirming that our advection-population dynamics
model is close to the random map when mixing is strong, we fix
$r_0=1/2$ and $r_1=4$, as in \cite{vulpio}. This describes a
system in which predators are advected over regions in which not
enough food is available ($r_0=1/2$ leads to population
extinction) and over the strip-like regions in which preys are
abundant (leading to chaotic population dynamics of the
predators). We iterate (\ref{lagran}) to obtain $C(i)\equiv
C(\bx(i),i)$, and then calculate the reaction Lyapunov exponent
$\lambda^R$ for our system:
\begin{equation}
\lambda^R=\lim_{N\to \infty}
\frac{1}{N}\sum_{i=0}^{N-1}\ln\left|G'\left(C(i)\right)\right|=
\lim_{N\to \infty}
\frac{1}{N}\sum_{i=0}^{N-1}\ln\left|\mu n\left(\bx(i)\right)
\left(1-2C(i)\right)\right|
\label{lyapunov},
\end{equation}
The initial condition $C(\bx,0)$ was a smooth function
proportional to $\sin(x)\sin(y)$. $\lambda^R$ measures the rate of
convergence or divergence of two initially similar concentration
values $C$ at the same fluid particle. In Fig.~\ref{fig:lambdar}
we show $\lambda^R$ as a function of $p$ for different values of
$K$. When $K$ is increased above a high enough value (e.g. $K
\approx 9$, for which the
standard map shows a unique ergodic chaotic region
\cite{Lichtenberg}), $\lambda^R$ approaches the Lyapunov exponent
$\lambda$ of the random map \cite{vulpio}. On the other side, when
$K$ is small, this correspondence is lost.

Therefore, exploiting this equivalence we can study different
regimes of our system depending on the value of the parameter $p$.
In particular, the spatial patterns of the advected field are
strongly dependent of the value of $\lambda^R$. Next section is
dedicated to the study of these structures.

\section{PREDATOR SPATIAL STRUCTURES}

The three regimes described above for the random map are also
found for the behavior of the Lyapunov exponent $\lambda^R$ as a
function of $p$ in our advection model, with just some minor
quantitative differences, e.g. in the values of $p_1$ and $p_2$
(in particular, in most of our calculations we take $K=9$, which
gives $p_1 \simeq 0.34 $ and $p_2 \simeq 0.48$). These three
regimes give rise to the following different predator spatial
structures:
\begin{itemize}
  \item[i)] For $p \leq p_1 $, the concentration of predators
vanishes in all the space. The nutrient area is too small to
support a stable population.
  \item[ii)] If $p_1 \leq p \leq p_2 $,
a typical spatial pattern appears. A relatively high, but very
intermittent, concentration of predators appears in the area
occupied by preys. Moreover, the concentration pattern displays
fractal features.
  \item[iii)] If $p \geq p_2  $,
the spatial concentration of predators shows a
random pattern. No typical structure seems to emerge.
\end{itemize}

Whereas the result for case i) is self-evident,  cases ii)
and iii) need a more elaborated study. We proceed in the following
subsections.

\subsection{Case $p_1 \leq p \leq p_2 $}

The regime which we have labeled  above with ii) is characterized
by a negative reaction Lyapunov exponent $\lambda^R$. In this
case, we observe numerically (see Fig. \ref{fig:smallp}) the
existence of a typical structure of the reactive field, which
follows the strip-like structure of the preys. Moreover, a fractal
pattern seems to be displayed by the distribution. Let us proceed
to a quantitative characterization of these features.

\begin{figure}
\centering
\epsfig{file=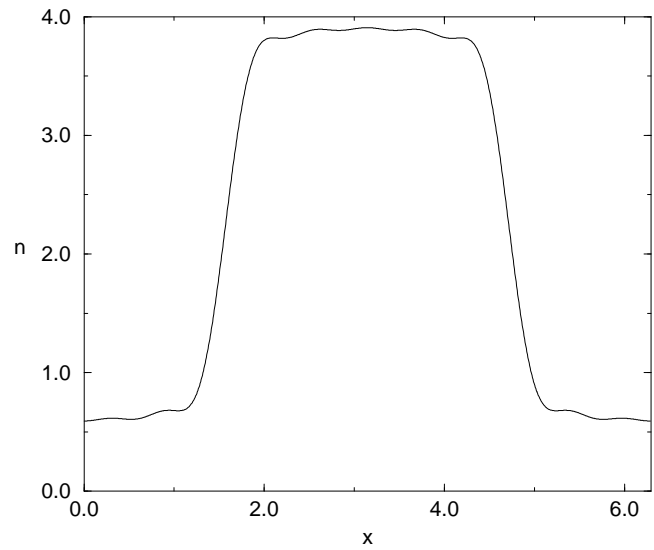,width=10cm}
 \caption{ Onedimensional cut of the continuous distribution
of nutrients. }
\label{fig:cut}
\end{figure}

References \cite{filam} and \cite{filam2} study the general
continuous-time case of a chemically or biologically decaying
field (thus with a negative reaction Lyapunov exponent) advected
by a chaotic 2d flow. Since periodic velocity fields are used, it
is straightforward to apply the results in these papers to our
case with discrete time. Nevertheless, a fundamental assumption in
these studies is that a source term in the equations, analogous to
our prey distribution, is a smooth function of space. The
quantitative results of \cite{filam,filam2} would fail (see e.g.
Eq.~(10) in \cite{filam}) when discontinuities are present in the
source, as in our localized prey distribution (\ref{step}).
Therefore, in our calculations, and with the view on
characterizing the spatial patterns, we will use a continuous
approximation to the previous step function describing the
distribution of nutrients, which would allow us to compare our
results (in the regime of $\lambda^R <0$) with those in
\cite{filam,filam2}. The approximation is performed by truncating
the Fourier transform of the step function of width $p$
(Eq.~(\ref{step}) and smoothing properly the coefficients
\cite{Canuto}. The final expression we use is:
\begin{equation}
n(x,y)= \frac{1}{2 \pi}\left(\pi + \frac{7 p}{2}\right) +
\sum_{k=-N, k
\ne 0}^{N} \left( 1-\frac{|k|}{N+1}\right) (-1)^k \frac{7}{2\pi
k}\sin\left(\frac{kp}{2}\right)\cos(kx),
\end{equation}
for any $0 \le x,y \le 2\pi$ and  $N=10$. Fig.~(\ref{fig:cut})
shows a $1d$ cut of this smoothed distribution of nutrients.

A quantitative characterization of the observed structures can be
performed in terms of structure functions. In particular, the
structure function of order one, $S_1$, is defined by:
\begin{equation}
 S_1({\bf \delta x}) = \left<\left|C(\bx+{\bf\delta x})-
 C(\bx)\right|\right>,
\end{equation}
where $<...>$ indicates an average taken over the different
spatial points $\bx$ along a line in the system. In \cite{filam2}
the scaling of $S_1$ is calculated with the result $S_1({\bf
\delta x})
\sim |{\bf
\delta x}|^\alpha$ when ${\bf \delta x}\to 0$, with $\alpha
\approx {\frac{|\lambda^R|}{\lambda^F}}$ when $\lambda^R < 0$
and $\lambda^F>|\lambda^R|$, being $\lambda^F$ the Lyapunov
exponent of the Lagrangian motion (\ref{velo}). The above
expression for $\alpha$ is just an approximation to which
multifractal corrections should be in principle added
\cite{filam2}, but we are not going to consider them here.

 The Lyapunov exponent of the standard map,
for $K$ high enough, is given \cite{Lichtenberg} by: $\lambda^F
\simeq \ln(K/2)$. In our calculations for $K=9$ we are in the above
mentioned conditions, that is, $\lambda^F>|\lambda^R|$ for all the
values of $p$.

\begin{figure}
\centering
\epsfig{file=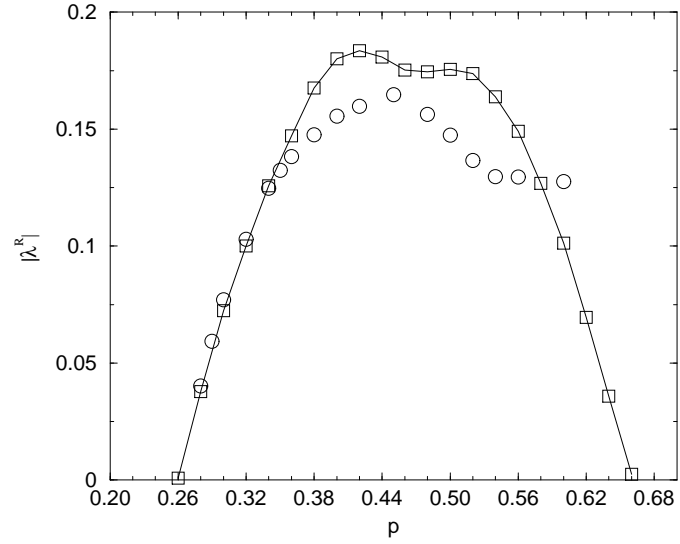,width=10cm}
\caption{ $|\lambda^R|$ for a continuous distribution of nutrients,
calculated in two different ways against $p$. Values labeled with
squares come from a direct calculation using expression
(\ref{lyapunov}) in the text. Circles are calculated from
$|\lambda^R|\approx \alpha \lambda^F$, being $\alpha$ the
numerically calculated scaling exponent of the first order
structure function. }
\label{fig:abslambdar}
\end{figure}

\begin{figure}
\centering
\epsfig{file=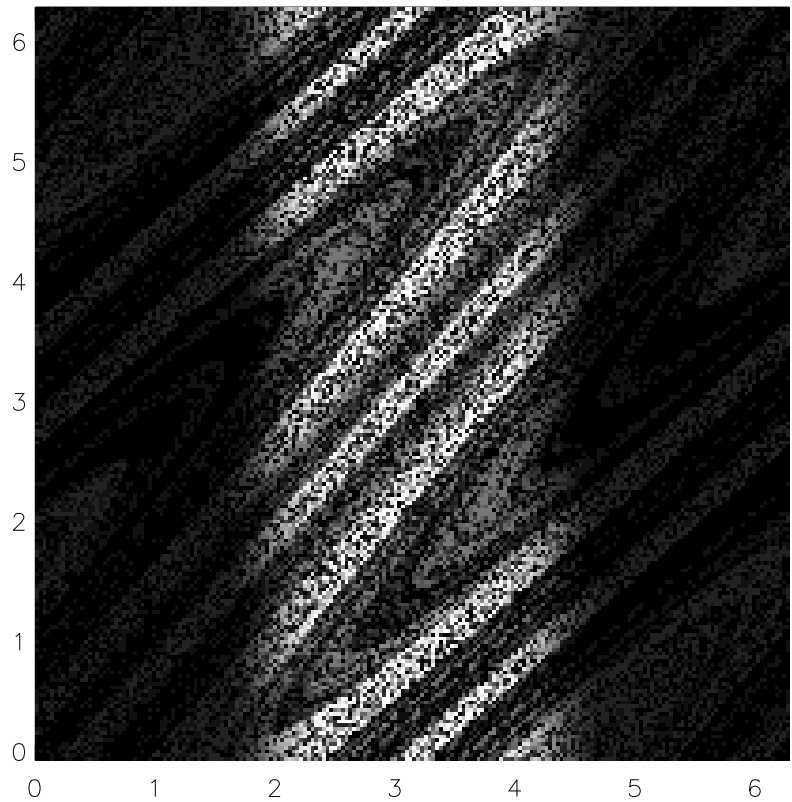,width=10cm}
\vspace{1cm}
\caption{Predator 2d pattern obtained for $p=0.37$.
The distribution of nutrients is continuous. The lighter the colour
the higher the concentration. }
\label{fig:smallp}
\end{figure}

Fig.~(\ref{fig:abslambdar}) shows $|\lambda^R|$ as a function of
$p$ for $K=9$ (the smooth approximation to the spatial
distribution of preys is used). In addition, we have numerically
calculated the scaling exponent $\alpha$ of $S_1$ in lines across
the central strip of nutrients, and multiplied it by
$\ln(9/2)\approx\lambda^F$ for different values of $p$. The
agreement between both quantities is quite good for $p$ near
$p_1$, confirming the expression $\alpha
\approx {\frac{|\lambda^R|}{\lambda^F}}$, although it gets worse
as $p
\to p_2$. The reason for this are the already mentioned
multifractal corrections to the scaling exponent of the structure
function, but a deeper discussion about this will be given in a
subsequent work. The agreement allows us to understand the pattern
displayed in Fig.~(\ref{fig:smallp}) in terms of the filamental
fractal patterns discussed in \cite{filam,filam2} for
continuous-time dynamics. The observed fractal structures are
revealing the stable and unstable manifolds (local contracting and
expanding directions) attached to each point of the phase space of
the standard map. It should be noted however that there is here a
much larger amount of irregularities at small scales than in the
patterns analyzed in \cite{filam,filam2}. The reason is the much
more irregular dynamics associated to the logistic map considered
here. The Lagrangian evolution of $C(t)\equiv C(\bx(t),t)$ is
related to the one of the logistic random map, which is of the
{\sl on-off intermittency} type. Models considered in
\cite{filam,filam2} displayed simple local relaxation behavior.
The small-scale structure seen in Fig.~\ref{fig:smallp} will
introduce stronger multifractal corrections in higher order
structure functions. We finally remark that when a discontinuous
distribution of preys such as (\ref{step}) is considered, the
relationship $\alpha
\approx {\frac{|\lambda^R|}{\lambda^F}}$ is not satisfied at all.


\subsection{Case $p>p_2$}

Now we proceed to study the case $p>p_2$. In this regime, the
reaction Lyapunov exponent $\lambda^R$ is positive, i.e. the
chemical or biological part of our system is also chaotic.

\begin{figure}
\centering
\epsfig{file=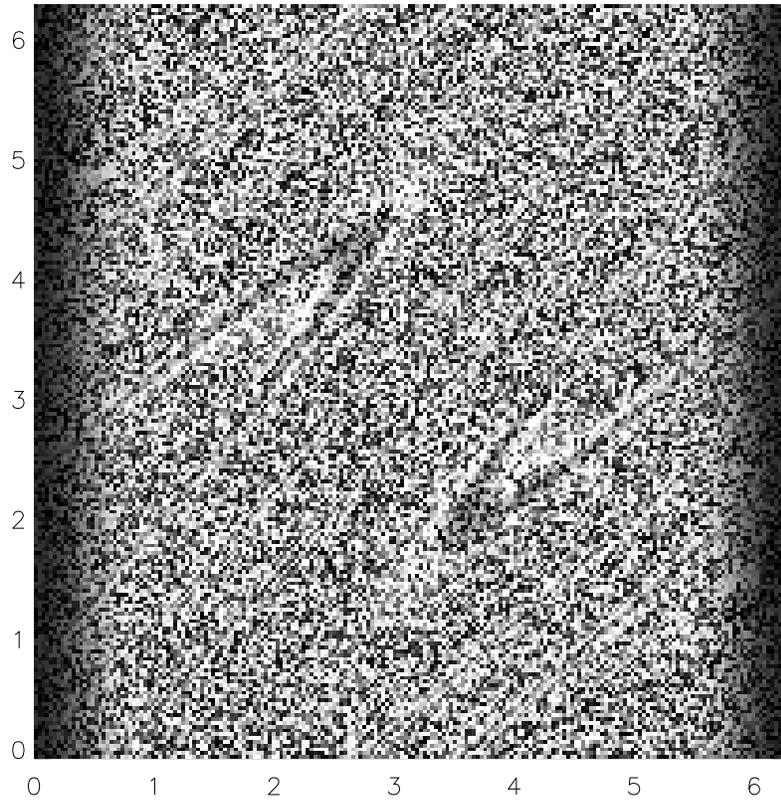,width=10cm}
\caption{Predator 2d pattern  obtained for $p=0.87$.
The distribution of nutrients is continuous. }
\label{fig:largep}
\end{figure}

The patterns calculated in this regime have random appearance,
being dominated by strong small-scale irregularity with very small
amount of structure (see Fig.~\ref{fig:largep}). In fact, the
scaling exponents of the first-order structure function are close
to zero, as corresponding to a random discontinuous field.

This random structure is easy to understand once one has realized that
$\lambda^R>0$ in this range of $p$: neighboring sites, even if
they have initially nearly similar concentration values, and even
when they remain close for long time so that they experience close
values of the sequence $\bx(t)$, will unavoidably develop growing
differences in in concentration values, thus leading to the
observed discontinuities at small scales.

\section{DISCUSSION}

Summing up, spatial structures with fractal features (of
filamental type) appear for the predator field in a range of
values of the size of the nutrient patch $p$. An increasing amount
of small-scale randomness appears when $p$ is increased, until
structure is finally lost. The analogy with the random map model
has allowed us to understand this behavior as being originated by
the change in the value of the reaction Lyapunov exponent
$\lambda^R$ when $p$ is varied. In particular, structure is lost
when $\lambda^R$ becomes positive. For $p$ small enough, global
extinction occurs, since most of the system has a parameter value
for which $C=0$ is the only attractor.

Our results have been obtained for a particular set of coupled
maps, and for a specific nutrient distribution. We do not expect
major qualitative changes in the above findings if the standard
map is replaced by a different advecting flow, as long as the
Lyapunov exponent $\lambda^F$ takes the same value. This belief is
supported by the more detailed arguments of Refs. \cite{filam} and
\cite{filam2} for the time-continuous case. It should be said
however that the quantitative strength of multifractal corrections
to simple expression such as $\alpha
\approx {\frac{|\lambda^R|}{\lambda^F}}$ will depend
on the particular flow chosen.

Our choice of the logistic map as the population dynamics to study
is certainly important for the results obtained. Our results
should describe the behavior under other population models as long
as their parameters take values favoring chaotic oscillations in a
localized portion of space, and favoring relaxation to a fixed
point in the rest. The election of the logistic map has allowed
the use of results known for random logistic maps, thus helping to
interpret the different patterns in terms of the value of
$\lambda^R$ and its relationship with $\lambda^F$. Those
quantities would be the right tool for the interpretation of
advection-reaction patterns in other population or chemical
models.

Diffusion has been discarded in the present work. We expect that
its only effect would be to smooth out any small-scale fractal or
random structure below a size of the order of
$\sqrt{D/\lambda^F}$. In fact, our numerical calculations have an
{\sl effective diffusion} which comes from our minimal spatial
resolution. As mentioned above, a more controlled way to introduce
diffusion is to perform explicitly, after each map operation, an
average of the concentrations of fluid particles closer than the
diffusion length.

We finally mention that the map approach turns out to be an
extremely efficient method from the numerical point of view, as
compared to direct solution of partial differential equations such
as (\ref{ard}) or other continuous approaches.

\section{Acknowledgments}
C.L. acknowledges a TAO (Transport processes in the Atmosphere and
Oceans) exchange grant of the European Science Foundation.
Financial support from CICyT (Spain) project MAR98-0840 is also
acknowledged.

\newpage

\newpage

\begin{center}
Figure 2 \\
L\'opez et al.
\end{center}

\vspace{2cm}

\newpage

\begin{center}
Figure 3 \\
L\'opez et al.
\end{center}

\vspace{2cm}

\newpage

\begin{center}
Figure 4 \\
L\'opez et al.
\end{center}

\vspace{2cm}

\newpage

\begin{center}
Figure 5 \\
L\'opez et al.
\end{center}
\vspace{2cm}

\end{document}